\documentclass[12pt]{iopart}
\usepackage{amssymb}
\usepackage{grffile}
\usepackage{graphicx}

\newcommand{\kk}{\mathbf{k}}
\newcommand{\I}{\mathrm{i}}
\renewcommand{\le}{\leqslant}

\begin{document}

\title[Magnetic States, Correlation Effects and Metal-Insulator Transition]{Magnetic States, Correlation Effects and Metal-Insulator Transition in FCC Lattice}

\author{M A Timirgazin$^1$, P. A. Igoshev$^{2,3}$, A. K. Arzhnikov$^1$ and V. Yu. Irkhin$^2$}

\address{$^1$Physical-Technical Institute, Ural Branch of Russian Academy of Sciences  - 426000 Izhevsk, Russia}
\address{$^2$Institute of Metal Physics, Ural Branch of Russian Academy of Sciences - 620990 Ekaterinburg, Russia}
\address{$^3$Ural Federal University - 620002 Ekaterinburg, Russia}
\ead{timirgazin@gmail.com}
\vspace{10pt}

\begin{abstract}
The ground-state magnetic phase diagram (including collinear and spiral states) of the single-band Hubbard model for the face-centered cubic lattice and related metal-insulator transition (MIT) are investigated within the slave-boson approach by Kotliar and Ruckenstein. 
The correlation-induced electron spectrum narrowing and a comparison with a generalized Hartree-Fock approximation allow one to estimate the strength of correlation effects.
This, as well as the MIT scenario, depends dramatically on the ratio of the next-nearest and nearest electron hopping integrals $t'/t$. 
In contrast with metallic state, possessing substantial band narrowing, insulator one is only weakly correlated.
The magnetic (Slater) scenario of MIT is found to be superior over the Mott one. 
Unlike simple and body-centered cubic lattices, MIT is the first order transition (discontinuous) for most $t'/t$. 
The insulator state is type-II or type-III antiferromagnet, and the metallic state is spin-spiral, collinear antiferromagnet or paramagnet depending on $t'/t$. 
The picture of magnetic ordering is compared with that in the standard localized-electron (Heisenberg) model.
\end{abstract}



\submitto{\JPCM}

\maketitle
%
%

\section{Introduction}

Magnetic materials with face-centered cubic (fcc) lattice are characterized by a great variety of magnetic states, \textit{e.\,g.}, antiferromagnets of type-I  (MnTe$_2$~\cite{Hastings59}); type-II (MnO~\cite{Bloch74}); type-III (MnS$_2$~\cite{Hastings59}); spiral (helical) magnetic structures ($\gamma$-Fe~\cite{gamma-Fe}). There is a number of complex compounds where magnetic moments form an effective fcc lattice, \textit{e.\,g.}, alkali fullerides A$_3$C$_{60}$ (A = K, Rb, Cs)~\cite{Capone09,Ganin10}, cluster compounds like GaTa$_4$Se$_8$, GaNb$_4$Se$_8$~\cite{Pocha05,Jakob07}, 'B site ordered' double perovskites~\cite{Battle83,Aczel13}. They demonstrate various magnetic orders as well and undergo metal-insulator transition (MIT) under pressure. The wealth of magnetic structures is largely explained by inherited geometric frustration of fcc lattice. The lattice can be decomposed into the system of edge-sharing tetrahedra, which makes impossible the simple N\'eel antiferromagnetic (AF) order~\cite{Lacroix11}.  

The frustrated magnetism in fcc structure is traditionally studied in the framework of the Heisenberg model which describes localized spins coupled by exchange interaction. If nearest-neighbor exchange integral $J$ only is taken into account,  the Fourier transformation of the exchange interaction
\begin{eqnarray}
J_\mathbf{Q}=
4J\left(\cos{\frac{Q_x}2}\cos{\frac{Q_y}2}+ \right.
\left.\cos{\frac{Q_y}2}\cos{\frac{Q_z}2}+\cos{\frac{Q_z}2}\cos{\frac{Q_x}2}\right)
\end{eqnarray}
approaches its minimum at 
$\mathbf{Q}=(0,q,2\pi)$ where $q$ is an arbitrary number in the $[0,\pi]$ range (lattice constant is taken as a unit), which determines the type of the ground state magnetic order being unspecified. For $q=0$ we have type-I AF structure, for $q=\pi$  type-III AF structure, the intermediate range corresponding to spiral (or spin-wave) structures which are, generally speaking, non-collinear and incommensurate. 
This ambiguity indicates the ground state degeneracy which is characteristic for  frustrated systems. 
Finite temperatures lift the degeneracy by favoring collinear magnetic structures owing to so-called ``order from disorder'' effect~\cite{Henley87,Canals04}. 

The account of the next-nearest-neighbor exchange $J'$ lifts the degeneracy already within the mean-field theory: 
depending on $J'$, one of the following collinear ground states is stable: type-I, type-II ($\mathbf{Q}=(\pi,\pi,\pi)$), type-III antiferromagnetic or ferromagnetic order~\cite{Lefmann01}. 
Monte--Carlo simulations predict type-I antiferromagnet at low temperatures for the nearest-neighbor model, while inclusion of  $J'$ should lead to type-III antiferromagnet which goes to type-I with increasing temperature~\cite{Gvozdikova05}. 
An account of magnetic anisotropy (\textit{e.g.}, Dzyaloshinsky-Moriya interaction) can stabilize type-I or type-III AF order~\cite{Heinila94}. 
In ref.~\cite{Khmelevskyi12}, Monte--Carlo simulations were performed for GdPtBi with exchange constants determined from first-principle calculations, and type-III antiferromagnet was found as the ground state. 

The localized magnetism picture is insufficient to describe properties of 
itinerant fcc magnets. 
In particular, the calculations based on the Heisenberg
model, as a rule, suppose exchange constants to be not changed with 
temperature, and can miss important effects: \textit{e.~g.} the results 
obtained for the Hubbard model on the square lattice show a significant 
change of the spiral magnetic order parameters with temperature~\cite{Arzhnikov13}. 

An alternative approach is provided by \textit{ab initio} band structure calculations of specific fcc materials based on the density functional theory with possible extensions like LDA+$U$ or 
LDA+DMFT which allow to take into account electron correlations. 
This approach enables one to describe  the experimentally observed ferromagnetic state in Ni and to explain peculiarities of its electron spectrum caused by correlation 
effects~\cite{Lichtenstein01,Braun06}. 
Modern band calculations reproduce successfully the experimental incommensurate spiral magnetic 
order in fcc $\gamma$-Fe revealing an important role of long-range 
competing exchange interactions~\cite{fcc-Fe}. 
Despite the considerable progress in predicting the properties of specific materials, band 
calculations lack generality and often cannot explain the physics 
leading to stabilization of one or another magnetic structure. 
To investigate the relative stability of all the possible commensurate and 
incommensurate magnetic states and to determine the role of correlations 
effects, depending on the system parameters, more general and fundamental approaches are needed.

The present paper is devoted to the construction of magnetic phase diagrams for the fcc lattice in the framework of Hubbard model which contains all essential physics of metallic magnetism and metal-insulator transition. In order to study spiral magnetism and electron correlations, we use the slave-boson approach (SBA)  proposed by Kotliar and Ruckenstein~\cite{Kotliar86} and generalized to spiral magnetic ordering~\cite{Fresard92}. 
In the saddle point approximation, SBA is qualitatively similar to the Gutzwiller approximation~\cite{Gutzwiller65}, the ground state energy being in a fair  agreement with the quantum Monte--Carlo and exact diagonalization calculations~\cite{Fresard92}.
To evaluate the impact of electron correlations on the formation of magnetically ordered states  and related metal-insulator transition we have used the Hartree--Fock approximation (HFA) as well. 
The HFA method yields qualitatively correct estimates for the energy of magnetically ordered phases even for large $U$, at least near half-filling~\cite{Irkhin89}. 
At the same time, in contrast to HFA, SBA yields a correct interpolation description of paramagnetic phase  energy too.

\section{Method}
\label{Formalism}
\subsection{Spiral magnetic state}

The SBA method was used to construct magnetic phase diagrams for the square, simple cubic and bcc lattices~\cite{Igoshev15JMMM}, providing reasonable results in a description of electron correlations and essentially improving the generalized HFA~\cite{Igoshev10}. 
Recent HFA and SBA were also applied to study magnetic and paramagnetic MIT's for bipartite square, simple cubic and bcc lattices~\cite{Timirgazin12,Timirgazin15}, importance of the spiral incommensurate magnetic states being demonstrated. 

We treat the single-band Hubbard model for the one-electron spectrum $t_{\kk}$,
\begin{equation}
      \label{eq:original_H}
      \mathcal{H}=\sum_{\kk\sigma\sigma'} t_\kk c^\dag_{\kk\sigma}c^{}_{\kk\sigma}+U\sum_i n_{i\uparrow}n_{i\downarrow}
\end{equation}
where $c^\dag_{\kk\sigma}$ ($c_{\kk\sigma}$) is a creation (annihilation) electron operator in one-electron state with quasimomentum $\kk$ and spin projection $\sigma$, and $U$ is Coulomb (Hubbard) interaction parameter, $n_{i\sigma}$ is particle number operator at site $i$. 
To treat  the spiral magnetic order formation we apply the generalized  SBA~\cite{Fresard92,Igoshev15JMMM,2013:Igoshev,Igoshev15JPCM} and directly minimize the ground state {\it grand potential} $\Omega(\mathbf{Q})$ with respect to spiral wave vector $\bf Q$. The calculations are based on the fermionic part of the effective Hamiltonian,
\begin{equation}
	\label{Hf}
	\mathcal{H}_{\rm f} = \sum_{\kk\sigma\sigma'} \left(z_{\sigma}z_{\sigma'}t_{\sigma\sigma'}(\kk)+\lambda_\sigma\delta_{\sigma\sigma'}\right)c^+_{\kk\sigma}c_{\kk\sigma'}
\end{equation}
where
\begin{equation}	
 \label{eq:t}
	t_{\sigma\sigma'}(\kk) = e^+_\kk \delta_{\sigma\sigma'} + e^-_\kk\delta_{\sigma,{-\sigma}'},
\end{equation}
\begin{equation}	
	e^\pm_{\kk} =(1/2)(t_{\kk+\mathbf{Q}/2}\pm t_{\kk-\mathbf{Q}/2}),
\end{equation}
\begin{equation}
	\label{z_def2}
	z_\sigma=(1-d^2-p_\sigma^2)^{-1/2} (ep_\sigma+p_{-\sigma}d) (1-e^2-p_{-\sigma}^2)^{-1/2} 
\end{equation}
is the correlation-induced one-electron band narrowing and $\lambda_\sigma$ is the one-electron shift. The average probability amplitudes of quantum many-electron states  (empty $e$, singly occupied $p_\sigma$, doubly occupied $d$) are calculated in a mean-field manner.
The corresponding eigenvalues (antiferromagnetic subbands) yielding the renormalization of the electron spectrum can be expressed analytically,
\begin{equation}
	\label{eq:subband_spectrum}
E_{s=\pm1}(\kk)=(1/2)\left((z^2_\uparrow+z^2_\downarrow)e^+_{\kk}+\lambda_\uparrow+\lambda_\downarrow\right)+ s\sqrt{D_\kk},
\end{equation}
where
\begin{equation}
    \label{eq:Dk}
    D_\kk=(1/4)\left((z^2_\uparrow-z^2_\downarrow)e^+_{\kk}+ \lambda_\uparrow-\lambda_\downarrow\right)^2+(e^-_{\kk}z_\uparrow z_\downarrow)^2.
\end{equation}
The bare electron spectrum in fcc lattice with account of nearest- and next-nearest neighbors hopping reads
\begin{eqnarray}
t_\kk = -4t\left(\cos{\frac{k_x}2}\cos{\frac{k_y}2}+\cos{\frac{k_y}2}\cos{\frac{k_z}2}+\cos{\frac{k_x}2}\cos{\frac{k_z}2}\right)+\nonumber\\
	+2t'(\cos{k_x}+\cos{k_y}+\cos{k_z}).
\end{eqnarray}
The electronic concentration $n$ and the magnetization $m$ satisfy mean-field-like equations \cite{Igoshev10} respectively,
\begin{equation}
	\label{eq:n_def}
	n=\frac1{N}\sum_{\kk s}f(E_s(\kk)),
\end{equation}
\begin{equation}
	\label{eq:m_def}
	m=\frac1{2N}\sum_{\kk s}sf(E_s(\kk))\frac{e^{+}_\kk(z^2_\uparrow-z^2_\downarrow)+ \lambda_\uparrow - \lambda_\downarrow }{\sqrt{D_\kk}},
\end{equation}
being in contrast with the former equations by that renormalization parameters $z^2_\sigma$ and $\lambda_\sigma$ are directly taken into account. 

The minimization of $\Omega(\mathbf{Q})$ with respect to $\bf Q$ was performed
numerically with $\textbf{Q}$
running in high-symmetry directions of the Brillouin zone: $(Q,Q,Q)$, $(Q,Q,2\pi)$, $(Q,2\pi,2\pi)$, $(0,0,Q)$, $(0,Q,Q)$ and $(0,Q,2\pi)$ directions were taken into account. The step of changing $Q$ was $0.05\pi$. The integration over the Brillouin zone was performed on the grid containing 70~$\kk$-points in each direction. 
The accuracy settings were adjusted, if necessary. Since the ground state $\Omega$   actually depends on the chemical potential
$\mu$ as a parameter, we can determine the dependence of the magnetic structure on $\mu$ taking into
account the possibility of the phase separation automatically~\cite{Igoshev10}, avoiding the tedious application of the Maxwell rule in the
negative compressibility regions.

SBA reproduces well-known Brinkman-Rice (BR) result~\cite{Brinkman70} for paramagnetic-state Mott transition. The consideration of magnetic states leads to a possibility of MIT by Slater scenario: a rise of  $U$ increases the magnetic moment and opens a gap in the electron spectrum at some critical $U_c$. For bipartite lattices with only nearest-neighbor electron hopping the magnetic MIT occurs at $U_c=0$, magnetic insulator state being N\'eel antiferromagnet. An account of next-nearest-neighbor hopping $t'$ results in finite values of $U_c$. 

In order to study the metal-insulator transition we fix the electron concentration $n=1$ adjusting $\mu$ correspondingly. 
In the insulating magnetically ordered state the chemical potential should be placed in the gap of one-electron spectrum $E_{\pm}(\kk)$, see Eq.~(\ref{eq:subband_spectrum}),
\begin{equation}
	\max_{\kk} E_{-}(\kk) < \mu < \min_{\kk}E_{+}(\kk).
\end{equation}
This implies that in our model any deviation from half-filling results in the shift of the Fermi level from the gap into one of subbands and, therefore, in impossibility of insulator state.  
To calculate the critical value of MIT $U_c$ we run over values of $U$ starting from zero until MIT is found.
In the case of second-order magnetic phase transition the determination of $U_c/t$ is not sufficiently precise (we assume $t>0$), and  we determine it by calculating the generalized magnetic susceptibility within SBA
\cite{1991:Li}.

\subsection{Antiferromagnetic  type-III state}

\begin{figure}
  \center
  \includegraphics[width=0.49\textwidth]{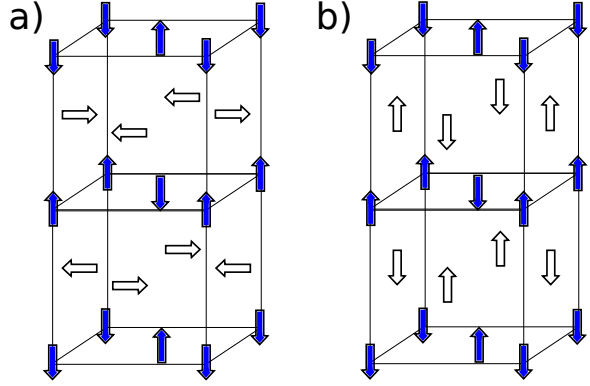}
  \caption{a) Non-collinear $(0,\pi,2\pi)$ spiral structure. b) Collinear type-III AF structure. Filled and empty arrows belong to the antiferromagnetic sublattices 'A' and 'B' respectively. }
	\label{fig:AF3}
  \end{figure}

While type-I and type-II AF structures can be viewed as spiral magnetic orders with $\mathbf{Q}=(0,0,2\pi)$ and $(\pi,\pi,\pi)$ correspondingly, for the type-III AF structure this is not possible. $(0,\pi,2\pi)$ wave vector, which is associated with type-III antiferromagnet, gives actually a {\it non-collinear} structure as a regular spiral  depicted in fig. \ref{fig:AF3}a. It can be divided into two sublattices ('A' and 'B'), each being ordered antiferromagnetically (filled and empty arrows in fig. \ref{fig:AF3}), so that their magnetic moments are perpendicular to each other. To obtain the type-III collinear order the magnetization direction of sublattice 'B' should be rotated by $\pi/2$ (fig. \ref{fig:AF3}b). This structure can be described by $\mathbf{Q}=(0,\pi,2\pi)$ as well, but its site magnetization dependence is specified by $m^x_j=m\cos{\left(\mathbf{QR}_j+\pi/4\right)}$, $m^y=0$ (that is a commensurate spin density wave), in contrast to spiral's one $m^x_j=m\cos{\mathbf{QR}_j}$, $m^y_j=m\sin{\mathbf{QR}_j}$. 
As for  experimental observations of the type-III AF phase (see Introduction), a collinear structure is usually implied. 
Experimental distinction between collinear type-III AF and non-collinear $(0,\pi,2\pi)$ orders is a non-trivial task which requires some special techniques such as polarized neutron diffraction on single crystals. 
The application of these measurements to the compound MnS$_2$ confirmed the formation of collinear type-III AF order at low temperatures~\cite{Chatto1991}.

The localized-moment picture predicts the same energy for the collinear type-III AF and non-collinear $(0,\pi,2\pi)$ states. 
This can be easily understood: total intersublattice interaction energy equals to zero due to antiferromagnetic structure and intrasublattice interaction energy does not depend on the relative orientation of sublattices. 
In the itinerant-electron systems these two states are not generally energetically degenerate 
and the calculations within the Hubbard model can identify the ground state;  such an investigation was not performed yet to our knowledge. 

For the sake of generality, we assume that the sublattice 'B' magnetization is rotated by some angle $\phi$ as compared to the trivial spiral order.   
Analogously to the Eqs.~(\ref{Hf}),(\ref{eq:t}) we write the fermionic part of the Hubbard Hamiltonian for the system of two sublattices,
\begin{eqnarray}
\label{eq:ham_phi}
\bar\mathcal{H}_{\rm f} = \sum_{\kk\sigma\sigma'\alpha} \left(z_{\sigma}z_{\sigma'}t^{\rm s}_{\sigma\sigma'}(\kk)+\lambda_\sigma\delta_{\sigma\sigma'}\right)c^+_{\kk\alpha\sigma}c_{\kk\alpha\sigma'}+ \nonumber\\
+\sum_{\kk\sigma\sigma'} \left(z_{\sigma}z_{\sigma'}t^{i}_{\sigma\sigma'}(\kk) c^+_{\kk A\sigma}c_{\kk B\sigma'}+{\rm h.c.}\right)
\end{eqnarray}
where $\alpha={\rm A,B}$ is a sublattice index; intrasublattice (s) and intersublattice (i) Fourier transforms of the hopping read
\begin{eqnarray}
t^{\rm s}_{\sigma\sigma'}(\kk) = e^+_{\kk,{\rm s}} \delta_{\sigma\sigma'} + e^-_{\kk,{\rm s}}\delta_{\sigma,{-\sigma}'},\nonumber \\
t^{\rm i}_{\sigma\sigma'}(\kk) = \left(e^+_{\kk,{\rm i}}\cos{\frac\phi2}+\I e^-_{\kk,{\rm i}}\sin{\frac\phi2}\right) \delta_{\sigma\sigma'} + \left(e^-_{\kk,{\rm i}}\cos{\frac\phi2}+\I e^+_{\kk,{\rm i}}\sin{\frac\phi2}\right)\delta_{\sigma,{-\sigma}'},\nonumber \\
e^\pm_{\kk,{\rm s(i)}} =(1/2)\left(t^{\rm s(i)}_{\kk+\mathbf{Q}/2}\pm t^{\rm s(i)}_{\kk-\mathbf{Q}/2}\right).
\end{eqnarray}
So $\phi=0$ and $\phi=\pi/2$ correspond to the spiral $(0,\pi,2\pi)$ and the collinear AF-III structures respectively. The bare electron spectra within ('s') and between ('i') sublattices have a form respectively,
\begin{eqnarray}
t^{\rm s}_\kk = -4t\cos{\frac{k_x}2}\cos{\frac{k_z}2}+2t'(\cos{k_x}+\cos{k_y}+\cos{k_z}),\\
t^{\rm i}_\kk = -4t\cos{\frac{k_y}2}\left(\cos{\frac{k_x}2}+\cos{\frac{k_z}2}\right).
\end{eqnarray}
Diagonalization of the Hamiltonian (\ref{eq:ham_phi}) with double number of $c$-operators can not be done analytically and requires numerical calculation of spectra in each $\kk$ point. 
Performing this procedure allows us to calculate the ground state thermodynamical potential $\Omega$ and include the collinear AF-III phase 
on the equal foot with a full set of spiral phases 
for the construction of the Hubbard model phase diagrams.

\begin{figure}[!h]
  \center
\includegraphics[width=0.49\textwidth]{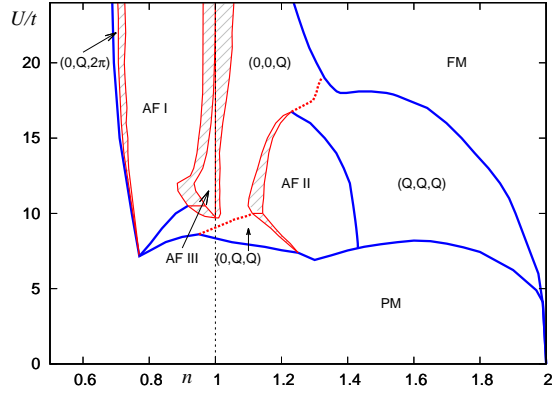}
\caption{
	(Color online)
	Ground state magnetic phase diagram of the Hubbard model for fcc lattice with $t'=0$ within SBA.
	The spiral phase regions are denoted according to the form of their wave vector (concrete {\it number} $Q$ depends on the point $(n,U)$ of the region).
	Filling shows the phase separation regions.
	Bold (blue) lines denote the second-order phase transitions.
	Solid (red) lines correspond to the boundaries between the regions of the homogeneous phase and phase
separation.
	Bold dashed (red) lines denote the first order phase transitions in the case where the region of the phase separation is narrow.
	Dashed (red) horizontal lines separate the phase separation regions corresponding to different phase pairs.
}
\label{fig:fcc_t'=0}
\end{figure}

\begin{figure}[!h]
  \center
\includegraphics[width=0.49\textwidth]{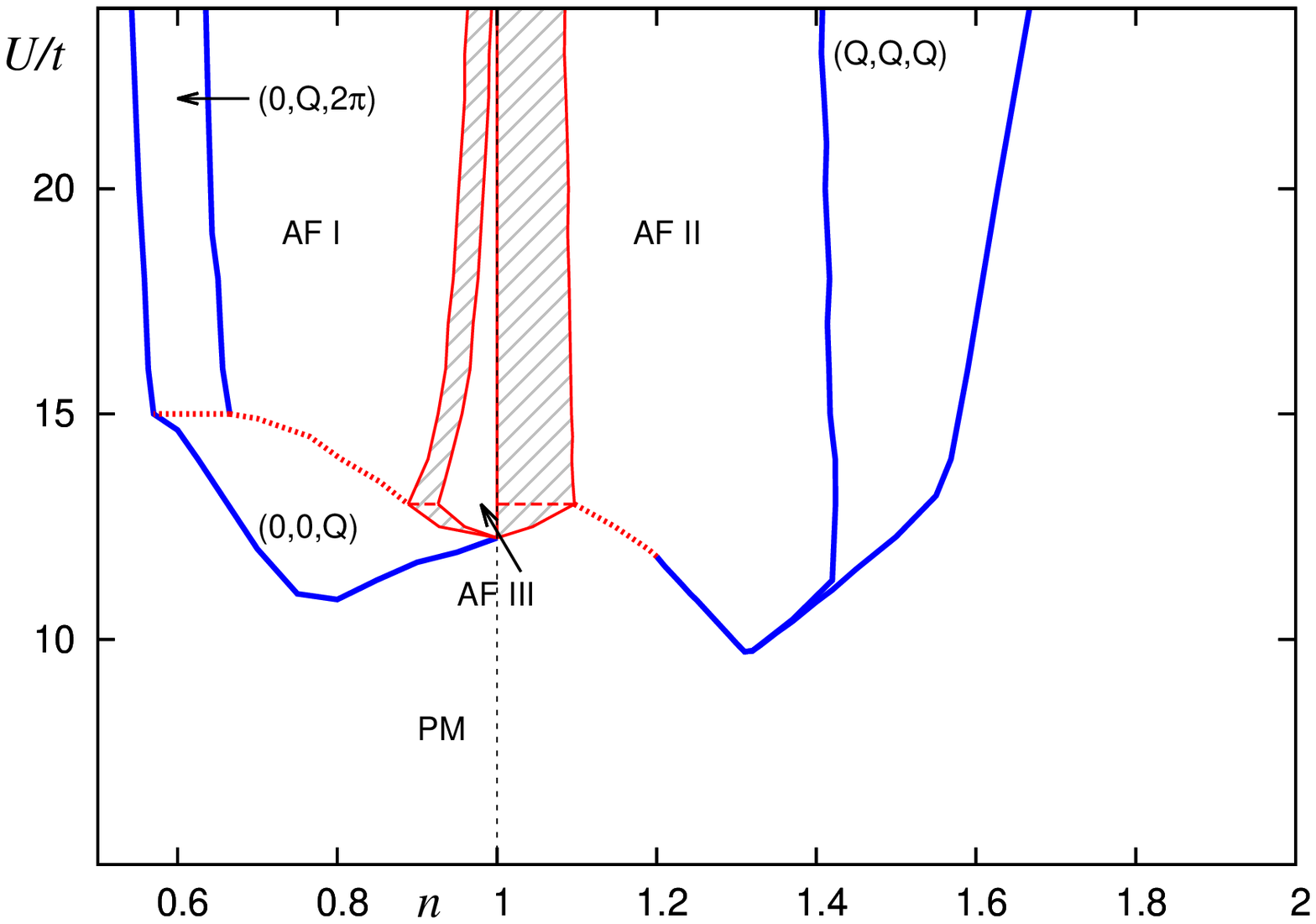}
\caption{
	(Color online)
	The magnetic phase diagram for fcc lattice with $t'=0.3t$.
	The notations are the same as in fig.~\ref{fig:fcc_t'=0}.
}
\label{fig:fcc_t'=+0.3}
\end{figure}

\begin{figure}[!h]
  \center
\includegraphics[width=0.49\textwidth]{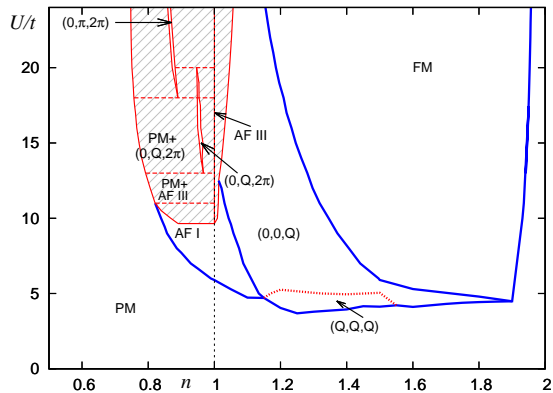}
\caption{
	(Color online)
	The magnetic phase diagram for fcc lattice with $t'=-0.3t$.
	The notations are the same as in fig.~\ref{fig:fcc_t'=0}.
}
\label{fig:fcc_t'=-0.3}
\end{figure}

\section{Results: magnetic phase diagrams}

We calculated the SBA phase diagrams of the ground state at following values of $t'/t$: $0$ (fig. \ref{fig:fcc_t'=0}), $0.3$ (fig. \ref{fig:fcc_t'=+0.3}) and $-0.3$ (fig. \ref{fig:fcc_t'=-0.3}). The diagrams are rich and contain the regions of paramagnetic, commensurate and incommensurate magnetic states. The commensurate phases include all three types of antiferromagnetic structures observed in real materials. The region of $(0,Q,2\pi)$ incommensurate spin-spiral order, observed in fcc-Fe~\cite{gamma-Fe}, is presented for all $t'/t$-values, being especially large for $t'=0.3t$. First-order transitions on the diagrams are accompanied by the phase separation regions. Similar diagrams and their detailed analysis were presented in ref.~\cite{Igoshev15JPCM}, but without including the type-III AF state. Here we focus on the differences owing to this additional phase. We see that for all considered values of $t'/t$ the AF-III phase becomes the ground state at half-filling and in its vicinity, replacing the non-collinear $(0,\pi,2\pi)$ phase (see figs. 10-12 in ref.~\cite{Igoshev15JPCM} for comparison). The only case where the $(0,\pi,2\pi)$-order survives is $t'=-0.3t$ one where the narrow region of the non-collinear state is present at $U/t>18$ and $n\approx0.87$ (fig. \ref{fig:fcc_t'=-0.3}).

\begin{figure}[!h]
  \center
\includegraphics[width=0.49\textwidth]{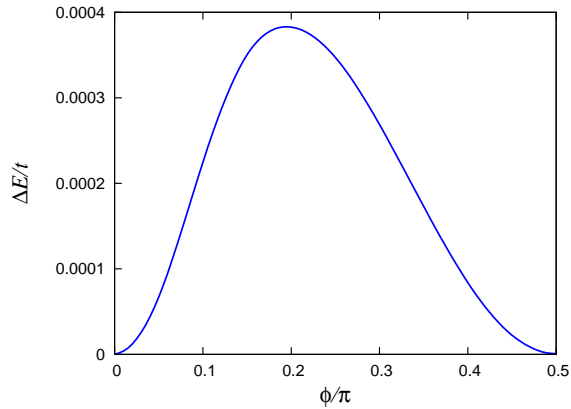}
\caption{
	The dependence of the total energy of the system on the angle between AF sublattices. The energy is measured from the minimal value $E=-0.5193911t$ corresponding to $\phi=0$ spiral order. The energy of type-III AF state ($\phi=\pi/2$) is almost equal and differs in 7-th decimal place.
}
\label{fig:phi}
\end{figure}

The wide area of the phase separation between $(0,\pi,2\pi)$ spiral state and type-III AF phase for $t'=-0.3t$ suggests a possibility of an intermediate magnetic order formation. While AF-III structure results from $(0,\pi,2\pi)$ order by rotating the AF sublattice 'B' by $\phi = \pi/2$ (fig. \ref{fig:AF3}), the general intermediate structure can be specified by an arbitrary angle $\phi$ between sublattices, see Eq.~(\ref{eq:ham_phi}). 
We choose a point $U=22t$, $n=0.96$ where the energies of pure $(0,\pi,2\pi)$ and AF-III phases are almost equal. For these parameters the total energy dependence on the angle $\phi$ between AF sublattices is calculated (fig. \ref{fig:phi}). The graph has two clear minima at $0$ and $\pi/2$ angles, so the intermediate angles turn out to be energetically unfavorable. An analogous behavior was found for other parameters studied. 
Note that extreme energy values differ on the fourth decimal place in the units of $t$. This means that all these magnetic states can be hardly distinguished between each other by limited in accuracy numerical simulation methods, like Monte--Carlo. Even if the required accuracy is reached the potential barrier between $(0,\pi,2\pi)$ and AF-III states does not allow to obtain $(0,\pi,2\pi)$ phase starting from the AF-III state and vice versa during simulation time. 

In ref.~\cite{Ulmke98} the Hubbard model on fcc lattice was studied in the dynamical mean-field theory taking into account the commensurate phases only. A finite temperature quantum Monte-Carlo method was used to calculate the phase diagrams extrapolated to $T=0$. No FM order was found for $t'=0$ and $U\leqslant6t$, which is in accordance with our diagram (fig.~\ref{fig:fcc_t'=0}) where the FM region for small values of $U/t$ is very narrow and can be hardly detected in finite temperature calculations. In the case  $t'=-t/4$ no magnetic order was found for $U=4t$ and the sequence of magnetic states AF$\rightarrow$PM$\rightarrow$FM$\rightarrow$PM was found for $U=6t$ starting from half-filling. This perfectly agrees with our diagram for $t'=-0.3t$ (fig.~\ref{fig:fcc_t'=-0.3}) except for the layer of the spiral $(0,0,Q)$-state between AF and FM regions (instead of PM layer), which was not considered in ref.~\cite{Ulmke98}. This qualitative agreement confirms the validity of the results obtained in this Section.

\begin{figure}[!h]
  \center
\includegraphics[width=0.49\textwidth]{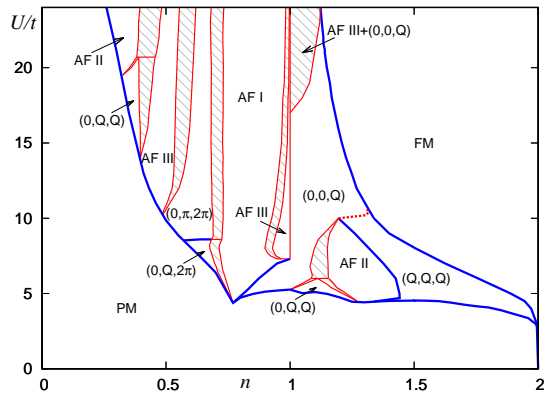}
\caption{
	(Color online)
	The magnetic phase diagram for fcc lattice with $t'=0$ in Hartree--Fock approximation.
	The notations are the same as in fig.~\ref{fig:fcc_t'=0}.
}
\label{fig:fcc_t'=0_HFA}
\end{figure}

To estimate the effect of electron correlations 
we present the analogous magnetic phase diagram at general $n$ for $t'=0$ using HFA (fig.~\ref{fig:fcc_t'=0_HFA}). 
(The particular case of MIT ($n = 1$) is discussed below, see Fig.~\ref{fig:MIT_HF}.) 
One can see that the HFA diagram turns out to be  considerably more rich and complicated than the SBA one, especially for $n<1$. Thus correlations result in reducing the variety of spiral magnetic phases. 
We have strong renormalization of FM phase boundaries at $n\lesssim 2$ by the cost of extension along $U$ axis of small $U$ spiral phase regions. 
On the other hand, at $n < 1$ the correct description of paramagnetic phase energy in SBA results in suppression of spiral magnetic phases lying in the vicinity of PM phase region boundary. 

\section{Results: metal-insulator transition}

An interesting feature of the diagrams \ref{fig:fcc_t'=0}--\ref{fig:fcc_t'=-0.3} is  the metal-insulator transition which occurs  in the half-filled band ($n=1$) due to opening the gap in electron spectrum with increasing $U$. For all three $t'$ values considered the insulator state appears to be the type-III AF one, the critical values $U_c$ of MIT being approximately $9.7t$, $12.2t$, $8.6t$ for $t'=0$, $0.3t$ and $-0.3t$ correspondingly. 
For a more detailed study of this important phenomenon we construct phase diagram of the model for $n=1$ in terms of $U$ and $t'$ variables.

\begin{figure}
  \center
  \includegraphics[width=0.49\textwidth]{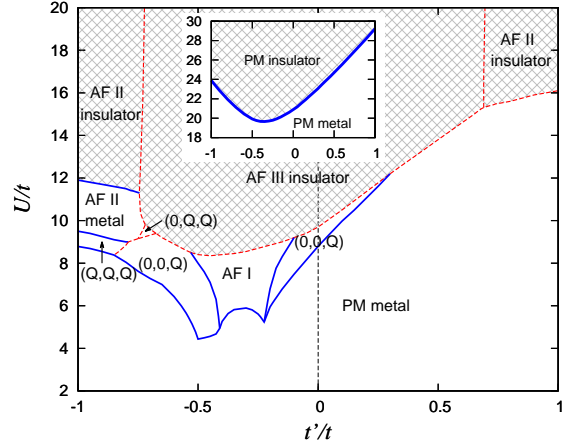}
  \caption{ (Color online)
	Phase diagram for fcc lattice at half-filling in terms of $U/t$ and $t'/t$ in SBA. 
  	Solid (blue) lines denote the second-order phase transitions. 
	Dashed (red) lines denote the first-order phase transitions. 
	Shaded (blank) area denotes the insulator (metal) state. 
	The inset shows the result of Mott scenario within the Brinkman--Rice criterion. 
	}
	\label{fig:MIT}
\end{figure}

Fig. \ref{fig:MIT} presents the phase diagram of the ground state at half-filling for $-t\le t' \le t$ with account of all the spiral states and the collinear type-III AF phase. 
We especially emphasize  the important  case of nearest-neighbor approximation ($t'=0$), discussed earlier in the Heisenberg model. 
While the inclusion of nonzero electron hopping integral $t$ (band picture) makes the collinear type-III AF order to be favorable among  other antiferromagnetic structures, the energy difference between type-I, collinear type-III and non-collinear ($0,\pi,2\pi$) orders tends to zero as $t$ decreases ($U/t\rightarrow\infty$).

\begin{figure}[!h]
  \center
  \includegraphics[width=0.49\textwidth]{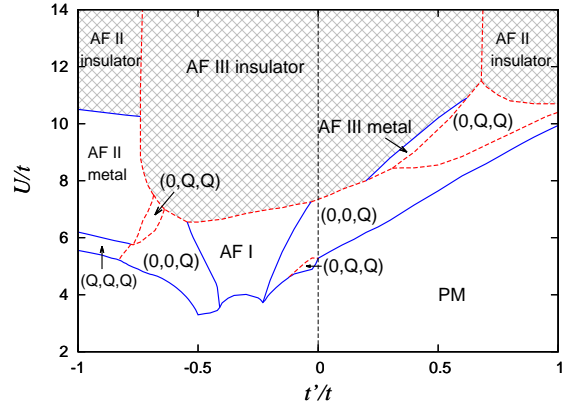}
  \caption{(Color online) Phase diagram for fcc lattice at half-filling in terms of $U/t$ and $t'/t$ in HFA.
  Notation is the same as in fig. \ref{fig:MIT}. }
	\label{fig:MIT_HF}
\end{figure}

The type-III AF phase remains the ground state in a wide range $-0.7t\lesssim t'\lesssim0.7t$ at large $U$.
This state is characterized by the presence of finite gap in electron spectrum and thus is the insulator. A decrease of $U$ leads to the transition into the gapless metallic state which is magnetic at $-0.7t\lesssim t'\lesssim 0.3t$ and paramagnetic at $0.3t\lesssim t'\lesssim 0.7t$. 
The most important result is that MIT is of first order (discontinuous), {\it e.g.}~the corresponding magnetic transition demonstrates discontinuous change of magnetic state. 
This differs fcc lattice from sc and bcc lattices where the second order MIT is found within the same model~\cite{Timirgazin12,Timirgazin15}. A possibility of the first order MIT in cubic lattices was discussed in ref.~\cite{Katsnelson84}, but concrete calculations were not performed yet.

The region of magnetic metallic state at $-0.7t\lesssim t'\lesssim 0.3t$ consists of two regions of vertical incommensurate spiral states $(0,0,Q)$, small pocket of $(0,Q,Q)$ spiral state and the type-I AF phase region with $Q=(0,0,2\pi)$. $(0,0,Q)$ state and the type-I AF phase transform into each other continuously through the second order phase transition. A decrease of $U$ leads to the second order transition to the paramagnetic (PM) phase.

The strong increase $|t'|\gtrsim0.7t$ changes the large-$U$ insulator phase from the type-III AF state to the type-II antiferromagnet with $\mathbf{Q}=(\pi,\pi,\pi)$. At positive $t'$, type-II AF phase goes to the PM metallic state with decreasing  $U$ through the first order transition. At negative $t'$ we have a completely different scenario: decrease of $U$ closes the electron spectrum gap and the type-II AF insulator smoothly goes to the type-II AF metal through the second order transition. Further on, type-II AF metal undergoes the second order transition to the $(Q,Q,Q)$-state.

The inset in fig. \ref{fig:MIT} presents the paramagnetic MIT obtained from the Brinkman--Rice criterion $U_{\rm BR}\equiv -16\sum_{\mathbf{k}}t_{\mathbf{k}}f(t_{\mathbf{k}})$, $f(\epsilon)$ being the Fermi function.  
Therefore we state that the Mott scenario is irrelevant at parameter values studied, since characteristic values of $U_{\rm BR}$ are much larger than those for magnetic MIT. 
The same conclusion is valid for the square, simple cubic  and bcc lattices~\cite{Timirgazin15}.

\begin{figure}
  \center
  \includegraphics[width=0.49\textwidth]{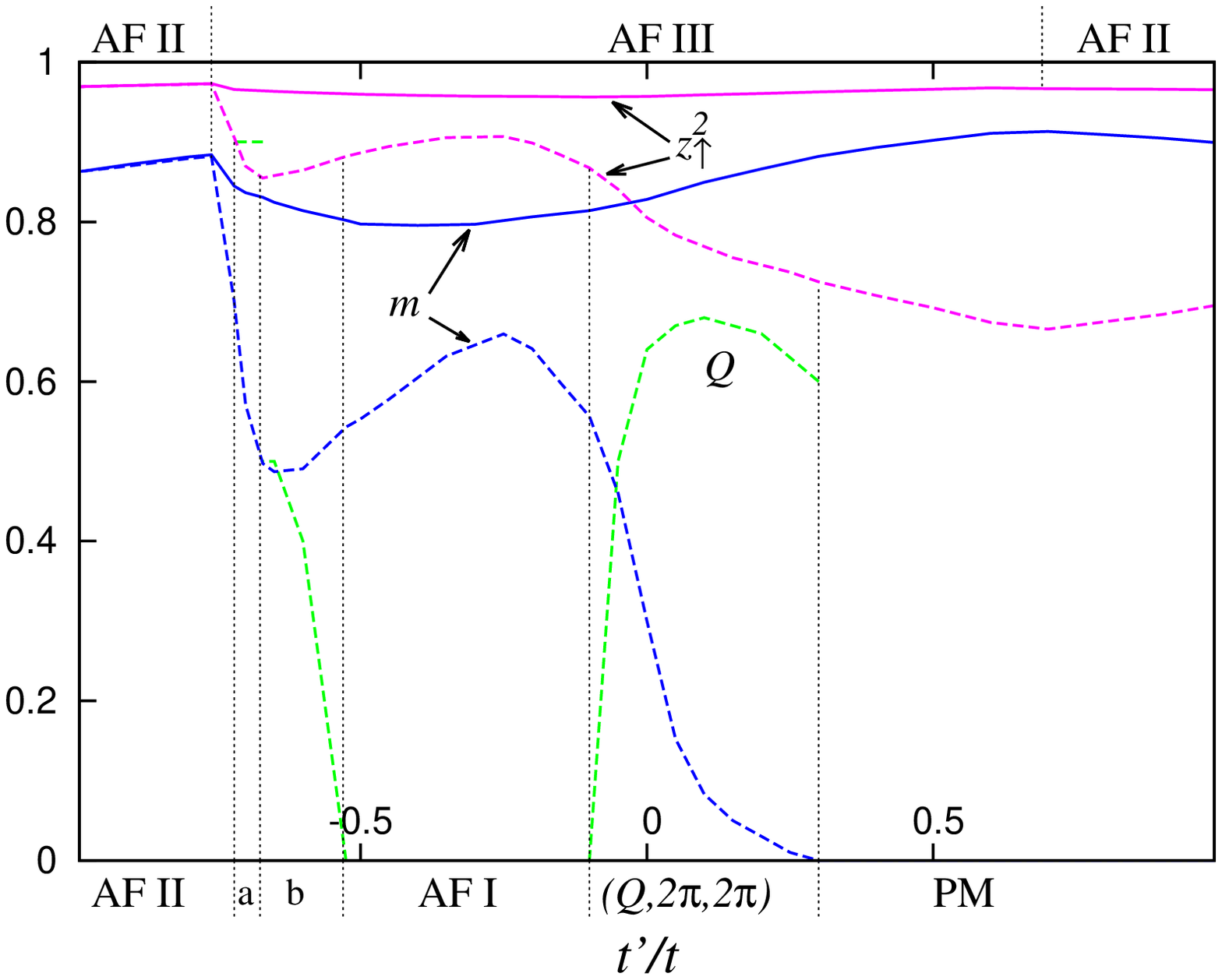}
   \includegraphics[width=0.49\textwidth]{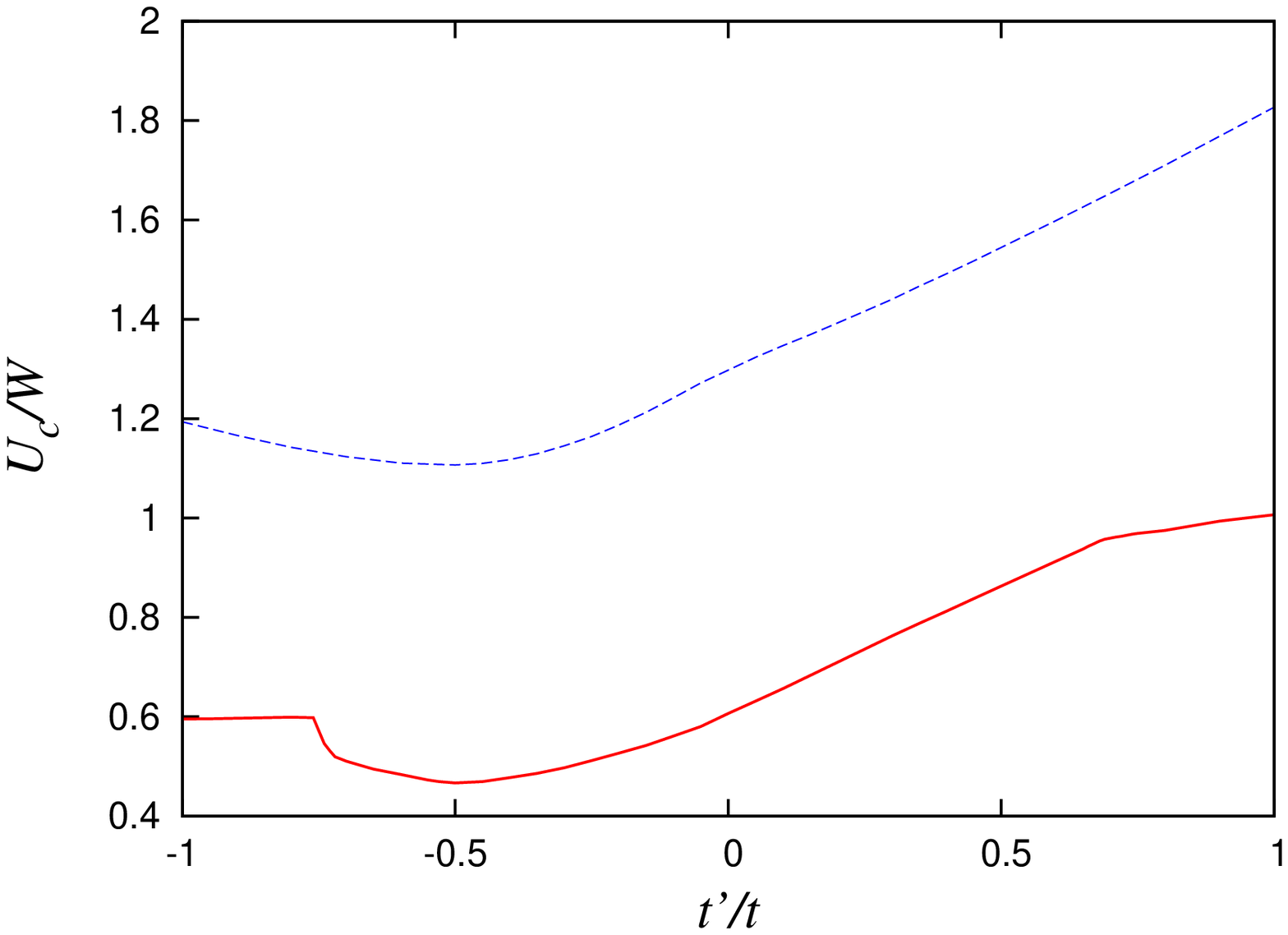}
  \caption{ (Color online)  
  (a) $t'/t$ dependence of physical quantities on both sides of the MIT boundary. Solid (dashed) green, blue and purple lines denote $Q/\pi$, the magnetic moment $m$ and $z^2_{\uparrow} = z^2_{\downarrow}$ factor of the insulator (metallic) state correspondingly. Vertical dotted lines separate the regions of different magnetic states. Upper (lower) labels denote the insulator (metal) magnetic states. Letters 'a' and 'b' denote the regions of $(0,Q,Q)$ and $(Q,2\pi,2\pi)$ spiral states correspondingly. $(Q,2\pi,2\pi)$ here is equivalent to $(0,0,2\pi-Q)$ due to the symmetry of electron spectrum.
  (b) The dependence of $U_c/W(t')$ vs $t'/t$ for the magnetic (solid line) and paramagnetic (dashed line) scenarios of MIT. 
  }
	\label{fig:aux}
\end{figure}

For comparison we constructed an analogous MIT phase diagram using HFA (fig. \ref{fig:MIT_HF}). In agreement with the conclusions of ref.~\cite{Igoshev15JMMM}, the main correlation effect is the renormalization of $U$, which results in a shift of magnetic regions to larger $U$'s within SBA. For negative $t'/t$ this is the only change of the HFA diagram as compared to SBA. For the positive-$t'$ half of the diagram the differences are more significant: the HFA diagram has additional region of spiral $(0,Q,Q)$ state; the metallic state is always magnetic at MIT; for $0.2t\lesssim t'\lesssim0.6t$ a narrow region of type-III AF metal phase exists within HFA going into type-III AF insulator through the second order transition. 
Physical reason of vanishing of metallic magnetic phase region below AF insulator and above paramagnetic metal regions for $t'\gtrsim0.3t$ within SBA is the following: from fig. 6 it is clear that metallic magnetic states are slightly more favorable within HFA than non-magnetic ones. Main shortcoming of HFA is strong overestimation of non-magnetic state energy due to artificial impossibility of reduction of the doubly occupied states number $d^2$.  
The only way to reduce the corresponding large Coulomb energy within HFA is inducing the magnetization~\cite{Igoshev15JPCM}.
Correct account of Coulomb energy of non-magnetic states within the SBA results in the suppression of magnetic metallic states. 
Note that the band narrowing $z^2$ for AF insulator states is close to 1 and its characteristics are close to HFA ones --- these states are almost not influenced by correlations effects, see details below (fig.~\ref{fig:aux}a). 

Analogous MIT phase diagrams for sc and bcc lattices~\cite{Timirgazin15} demonstrate tending of $U_c$ to zero at $t'\rightarrow 0$. In fcc lattice $U_c$ is always finite because this is not bipartite lattice. MIT in sc and bcc lattices is a continuous second order transition without a change of the magnetic state which is $(\pi,\pi,\pi)$ N\'eel antiferromagnet (type-II AF). In contrast, MIT in fcc lattice is of the first order in most of the studied $t'$ range, except for the region of large negative $t'$, where we have a picture similar to other cubic lattices, \textit{i.e.} continuous MIT without change of magnetic order. This can be interpreted as follows: at large absolute values of $t'$ the electron spectrum of fcc lattice tends to the the spectrum of sc lattice~\cite{Loly72}, and this tendency results in a similar magnetic state with $\mathbf{Q}=(\pi,\pi,\pi)$ and similar scenario of MIT. According to this analogy one may expect the $(\pi,\pi,\pi)$ metal region at large positive values of $t'/t$, but  magnetism in this region is possible only at very high  $U$ values where existence of the gapless magnetic state is impossible.

Fig. \ref{fig:aux}a illustrates the $t'/t$ dependences of the metal and insulator characteristics at the MIT boundary. The jump of magnetic moment at MIT from a magnetic to the paramagnetic phase ($t'/t\gtrsim0.3$)  is particularly large  and reaches 0.9 in $\mu_B$ units. 
If MIT occurs between two magnetic states the jump is also appreciable and mostly varies from 0.2 to 0.4. For the insulator state the $z^2$-factor is close to 1 in the whole  range of $t'/t$ which implies weakness of correlation effects. This agrees with the conclusion of paper~\cite{Igoshev15JPCM} about small role of correlation effects in vicinity of half-filling. At the same time for metallic state $z^2$ is noticeably smaller, especially for positive $t'$, so that  
the electron correlations are more significant for $t'>0$. 
Fig. \ref{fig:aux}b shows the critical $U_c$ Coulomb interaction parameter in units of bandwidth $W$ for the magnetic and paramagnetic (Brinkman--Rice) scenarios of MIT.
Note that $W$ depends on the $t'$ value if it has negative sign ($W=16t-4t'$), while for positive $t'$ the dependence is absent ($W=16t$).
It is clearly seen that the magnetic MIT occurs at approximately two times smaller values of $U_c$ than the paramagnetic MIT in all the range of $t'/t$. Interesting that the shape of both curves is very similar, indicating that $t'/t$-dependence of correlation effects is qualitatively the same for both magnetic and paramagnetic phases.
It should be noted that $U_c/W$ at $t\lesssim -0.7t$ in AFM-II phase is almost independent on $t'/t$, which implies that system goes into a different regime. 
This agrees with that the phase diagram for $t'=0.3t$ (fig.~\ref{fig:fcc_t'=+0.3}) is more strongly modified by correlations (as compared to the $t'=0$ diagram, fig.~\ref{fig:fcc_t'=0}) than $t'=-0.3t$ diagram, fig. \ref{fig:fcc_t'=-0.3}. One can observe the rapid growth of $U_c$ (in units of $W$, fig. \ref{fig:aux}b) simultaneously with considerable suppression of $z^2$-factor in metallic state (fig.~\ref{fig:aux}a), which can be interpreted as an enhancement of correlation effects for $t'>0$.

A quantitative comparison with the real compounds (see the Introduction) is hardly possible because of simplicity of our model which does not account for multiple bands, spin-orbit effects etc. However, a qualitative connection with some results for fullerides can be established. Cs$_3$C$_{60}$ is AF insulator at ambient conditions which becomes superconducting under pressure. The lowest unoccupied C$_{60}$ molecular orbitals are threefold degenerate, but Jahn--Teller distortion lifts the degeneracy by pairing two electrons, only one electron remaining unpaired~\cite{Zadik15}. Therefore the system can be roughly described by a single-band model , and we can state an agreement between the observed first-order transition from AF insulator to PM superconductor~\cite{Ihara10} with the first-order transition from AF insulator to PM metal at $t'\gtrsim0.3t$ in fig. \ref{fig:MIT}. A qualitative analogy can be carried out with MnO, where the nearest and next-nearest exchange integrals are of the same order~\cite{Lines65} and a similar pressure induced first-order transition from type-II AF insulator to PM metal occurs~\cite{Yoo05}. It should be noted that the metal-insulator transition in MnO is accompanied by a structural change which does not allow a direct comparison.
While generally first-order transitions imply a possibility of magnetic phase separation, no evidences of this were found for Cs$_3$C$_{60}$~\cite{Zadik15}. 
This result is confirmed by our calculations and is not surprising because the separation on underdoped and overdoped phases is practically impossible in the presence of the gap.

\section{Conclusions}

To conclude, we found that the account of conduction electron degrees of freedom allows to catch some interesting physical features of magnetic states which are discarded within the localized-electron model. 
The electron transfer effects lift the degeneracy of magnetic states ({\it e.~g.}, of collinear and spiral AF type-III  ordering) which is present within the localized model. 
At large $U/t$ the Anderson kinetic exchange mechanism establishes the equivalence of the Heisenberg and half-filled Hubbard models, and the transition from AF-III to AF-II occurs at $|t'|/t\sim 0.7$ (fig.~\ref{fig:MIT}),  which is in excellent agreement with the corresponding result of ref.~\cite{Lefmann01}, $J' =J/2$. 
However, at smaller $U/t$ the phase diagram turns out to be strongly asymmetric with respect to the sign of $t'$ and demonstrates 
presence of antiferromagnetic and spiral metallic phases between paramagnetic metal and antiferromagnetic insulator  regions. These metallic phases are influenced differently by correlations effects: at $t' > 0$ these states appears to be unstable due to bandwidth suppression and the magnetic order vanishes for most cases; at $t' < 0$ these regions are stable at the cost of substantial reduction of magnetization amplitude. 

\ack
We are grateful to A.N.~Ignatenko for useful discussions. 
The research was carried out within the state assignment of FASO of Russia (theme ``Quantum'' No. 01201463332 and Financial Program AAAA-A16-116021010082-8). 
This work was supported in part by the Division of Physical Sciences and Ural Branch, Russian Academy of Sciences
(projects no. 15-8-2-9, 15-8-2-12), by the Russian Foundation for Basic Research (projects no.
16-02-00995, 16-42-180516) and Act 211 Government of the Russian Federation 02.A03.21.0006.
The main amount of calculations was performed using the ``Uran'' cluster of IMM UB RAS.

\end{document}